\documentclass[preprint2]{aastex631}

\usepackage{amsmath}
\usepackage{tabularx}
\usepackage{mathrsfs}
\usepackage{xcolor}
\usepackage{CJK}
\usepackage[caption=false]{subfig}

\shorttitle{Primordial Alignment}
\shortauthors{Huang et al.}

\graphicspath{{./}{figures/}}

\begin{document}

\begin{CJK*}{UTF8}{gbsn}
\title{Primordial Orbital Alignment of Sednoids}

\author[0000-0003-1215-4130]{Yukun Huang (黄宇坤)}
\affiliation{Dept.~of Physics and Astronomy, University of British Columbia, 
6224 Agricultural Road, Vancouver, BC V6T 1Z1, Canada}
\affiliation{Department of Astronomy, Tsinghua University, Beijing 100084, China}
\author[0000-0002-0283-2260]{Brett Gladman}
\affiliation{Dept.~of Physics and Astronomy, University of British Columbia, 
6224 Agricultural Road, Vancouver, BC V6T 1Z1, Canada}

\begin{abstract}
We examined the past history of the three most detached TransNeptunian Objects (TNOs) -- Sedna, 2012 VP$_{113}$, and Leleakuhonua (2015 TG$_{387}$) -- the three clearest members of the dynamical class known as sednoids, with high perihelia distances $q$. 
By integrating backward their nominal (and a set of cloned) orbits for the Solar System's age, we surprisingly find that the only time all their apsidal lines tightly cluster was 4.5~Gyr ago, at perihelion longitude $\varpi$ of $200^\circ$. 
This ``primordial alignment'' is independent of the observational biases that contribute to the current on-sky clustering in the large-semimajor axis Kuiper Belt. 
If future sednoid discoveries confirm these findings, this strongly argues for an initial event during the planet formation epoch which imprinted this particular apsidal orientation on the early detached TNO population. Their apsidal orientations were then subsequently modified only by the simple precession from the 4 giant planets (and weakly by the galactic tide). 
If other sednoids also cluster around the same primordial value, various models suggesting a still present planet in the outer Solar System would be incompatible with this alignment. 
We inspected two scenarios that could potentially explain the primordial alignment. 
First, a rogue planet model (where another massive planet raises perihelia near its own longitude
until ejection) naturally produces this signature. 
Alternatively, a close stellar passage early in Solar System history raises perihelia, but it
is poor at creating strong apsidal clustering. 
We show that all other known $35<q<55$~au TNOs are either too perturbed or orbits are still too uncertain
to provide evidence for or against this paradigm.
\end{abstract}

\keywords{Trans-Neptunian objects (1705) ---  Kuiper belt (893) --- Celestial Mechanics (221) }

\section{Introduction} \label{sec:intro}
\end{CJK*}

\begin{table*}[htp]
  \caption{Barycentric orbital elements of the 
 three sednoids}
  \begin{center}
  \tabcolsep5pt
  \begin{tabular*}{1.0\textwidth}{l @{\extracolsep{\fill}} rrrrr}
  \hline\hline
      Object & $a$ (au) & $q$ (au) & $i$ ($\deg$) & $\Omega$ ($\deg$) & $\varpi$ ($\deg$) \\ 
      \hline
      2012 VP$_{113}$ & $262.0 \pm 0.6$ & 80.5 & 24.1 & 90.8 &24.7  \\
      (90377) Sedna & $506.4 \pm 0.2$ & 76.2 & 11.9 &144.4 &95.7 \\
      (541132) Leleakuhonua & $1089.6 \pm 185$ & 65.0 & 11.7 & 301.0 & 59.0  \\
      \hline
  \end{tabular*}
  \end{center}
  {\bf Note.} Data retrieved from JPL Small-Body Database (\url{https://ssd.jpl.nasa. gov/tools/sbdb_query.html}). Only the  semimajor axis uncertainties (1$\sigma$) 
  are presented.  Uncertainties in other orbital elements are too small to be listed. Compared to the other two objects, (541132) Leleakuhonua has a significantly large uncertainty in $a$, due to fewer observations and a shorter data-arc span.
  \label{tab:sednoids}
\end{table*}

The vast extent of the Solar System's Trans-Neptunian Objects (TNOs) has long captivated the curiosity of astronomers. These icy remnants, relics from the early Solar System, offer invaluable insights into the primordial conditions and dynamical histories of the giant planets. Among thousands of discovered TNOs, a tiny subset known as sednoids -- characterized by their large semimajor axes ($a$) and significantly high perihelia ($q$) -- stands out as particularly intriguing. The first member, (90377) Sedna \citep{Brown.2004}, was followed by 2012 VP$_{113}$ \citep{Trujillo.2014}, and most recently, (541132) Leleakuhonua (provisional designation: 2015 TG$_{387}$), was discovered by \citet{Sheppard.2019}. The barycentric orbital elements of the three sednoids are listed in Table~\ref{tab:sednoids}, in which the longitude of perihelion is defined as $\varpi = \Omega + \omega$.

Non-classical TNOs (see \citealt{Gladman.2021} for TNO classifications) are generally believed to have originated in the primordial planetesimal disk interior to $\sim$30~au and were scattered and/or transported onto their current orbits. It is believed that sednoids were created in a similar way due to a combination of early planetary scatterings and a detachment process that significantly increased their perihelia. The postulated $q$-raising mechanisms include: stellar flybys while the Sun was still in its birth cluster \citep{Morbidelli.2004,Kenyon.2004,Brasser.2006,Brasser.2012}, short-lived rogue planets \citep{Gladman.2006}, a distant planetary-mass solar companion \citep{Gomes.2006} or a still existing planet \citep{Lykawka.2008, Batygin.20167us}, as well as stellar flybys during solar migration in the Milky Way \citep{Kaib.2011}. 

Upon the discovery of 2012 VP$_{113}$, \citet{Trujillo.2014} pointed out that argument of perihelion ($\omega$) for many large-$a$ TNOs may cluster about $0^\circ$ and proposed a hypothetical super-Earth planet as an explanation. \citet{Batygin.20167us} explored a similar idea and proposed the so-called ``Planet Nine'' to account for the current on-sky orbital clustering of some distant TNOs. However, modern outer Solar System surveys \citep{Shankman.2017cuc, Napier.2021, Bernardinelli.2022} have cast doubt on the validity of the {\it current} clustering, 
showing it to be consistent with survey biases applied to
underlying uniform distributions in $\Omega$, $\varpi$, and $\omega$.

In this Letter, we presented a potentially new phenomenon involving the three sednoids that could provide a valuable constraint on their origin and the early history of the outer Solar System.

\begin{figure*}[htb!]
  \centering
  \includegraphics[width=0.95\textwidth]{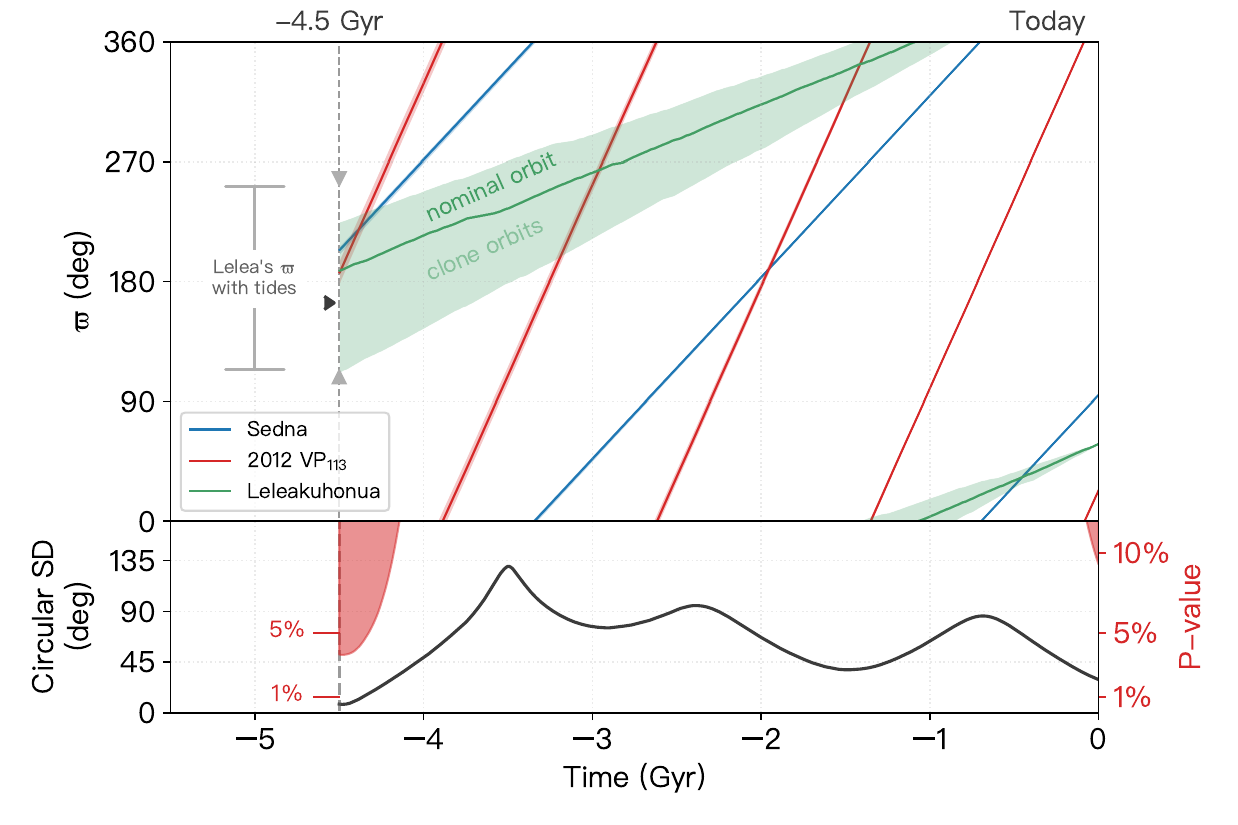}
  \caption{Upper: Past evolutions of perihelion longitudes ($\varpi$) for Sedna (blue), 2012 VP$_{113}$ (red), and Leleakuhonua (green). Solid lines denote the backward propagation of their nominal orbits without the galactic tide, whereas shaded areas denote ranges of their clone orbits. The only instance where the sednoid apsidal lines all converged was around 4.5 Gyr ago (vertical dashed line), right after the Solar System was formed. Leleakuhonua's primordial $\varpi$ and its range considering the galactic tide are marked by the black and grey triangles, respectively. Lower: The circular standard deviation of the three angles (black) and the statistical confidence ($p$-value, red shaded) that they are generated from a uniform distribution. There is only 1 in 20 chance (2$\sigma$) that three random angles would cluster at the same level as the sednoids did 4.5~Gyr ago. }
  \label{fig:sednoids-rewind}
\end{figure*}

\section{Rewinding Sednoids}\label{sub:rewinding}

\subsection{Analytical prediction}
The orbital evolutions of sednoids over the past $\approx$4~Gyr are primarily driven by secular precessions induced by gravitational effects of the four giant planets, assuming that no planetary mass is still present in the outer Solar System. The analytically approximate apsidal precession rate as a function of the TNO's ($a,e,i$) orbital elements is given by \citep{Batygin.2019}:
\begin{equation}\label{eq:apsidal-precession-rate}
    \begin{aligned}
        \dot{\varpi} = \frac{3}{8} \; n \;  \frac{3\cos^2{i}-1}{\left(1-e^2\right)^2} \sum_{j=5}^{8} \frac{m_j a_j^2}{M_\odot a^2}\;,
    \end{aligned}
\end{equation}
where $n$ denotes the TNO's mean motion, $M_\odot$ is the solar mass, and the index $j$ denotes the $j$-th planet. This equation estimates the apsidal precession periods of 2012 VP$_{113}$ (1.2~Gyr), Sedna (2.8~Gyr), and Leleakuhonua (7.1~Gyr). Assuming that their orbital $a$, $q$, and $i$ have not changed significantly post formation, one can rewind their longitudes of perihelion back in time by applying the linear precession rate (Equation~\ref{eq:apsidal-precession-rate}). 
Surprisingly, we found that the only time their $\varpi$ were all tightly clustered was 4 -- 4.5~Gyr ago, at 
around $\varpi\approx200^\circ$. 

\subsection{Backward integration}
Intrigued by this approximate analytical result, we carried out backward numerical integration to validate the primordial alignment of the sednoids. 
For each sednoid, we generated 11 initial conditions, including 1 nominal orbit and 10 cloned orbits distributed within the orbital uncertainty, using \textsc{SBDynT} 
\footnote{Small Body Dynamics Tool developed by Kat Volk 
and Dallin Spencer, 
\url{https://github.com/small-body-dynamics/SBDynT}},
with the clones serving to diagnose plausible uncertainties in the apsidal angles.
These initial test-particle orbits along with the four giant planets were integrated backward in time
using the \textsc{Mercurius} integrator in \textsc{Rebound} \citep{Rein.2012, Rein.2019}.  

Figure~\ref{fig:sednoids-rewind} shows the past evolution histories computed of $\varpi$. Similarly to the analytical approximation, their apsidal lines were aligned at 200$^\circ$ 4.5~Gyr ago, with a circular standard deviation of only $8^\circ$ (black curve in the lower panel). For reference, three randomly-generated angles would have an average circular standard deviation of $\approx$$60^\circ$.

To test the statistical significance of the clustering, we applied a Rayleigh test of uniformity at all the output times. 
The $p$-value of the test (red shaded curve and right-hand scale in the lower panel of 
Figure~\ref{fig:sednoids-rewind}) 
represents the probability that three random angles are more clustered than the three sednoids at that time. 
The Rayleigh test shows
that there is a $<$5\% chance that the primordial clustering is just a statistical coincidence. 
In particular, it would seem in principle that the alignment, if it were by chance, could occur at any time in the past; the fact that it occurs at the `special' time that corresponds to the planet formation epoch of Solar System history is a `difficult to quantify' additional low-probability event.

\subsection{Galactic Tide}
The galactic disk creates additional
dynamics for large-$a$ bodies via tidal forces, inducing $q$ and $i$ oscillations and an apsidal precession around the galactic pole \citep{Heisler.1986}. 
The galactic tidal effect becomes strong beyond $\sim$2000~au, where the tidal torquing timescale is comparable to the planetary scattering timescale \citep{Duncan.1987},
and thus should be tiny to small for the three sednoids. 
\citet{Sheppard.2019} integrated forward the three sednoids considering the galactic tide and the four giant planets, and found that the galactic tide has almost no effect on Sedna and 2012 VP$_{113}$, but does create a small $\pm6$~au oscillation of Leleakuhonua's perihelion (their figure 7); 
the authors concluded that Leleakuhonua is stable to the galactic tide, but do not show its $\varpi$ evolution with the tide.

To verify whether adding the tide would 
alter the $-4.5$~Gyr alignment,
we back-integrated the same nominal and clone orbits, considering both giant planets and the galactic tide. 
Due to its dominance,
we modeled only the vertical component of the tide \citep{Heisler.1986}, using the standard local density of the galactic disk of 0.1 $M_\odot$/pc$^3$ \citep{Levison.2006}. 
We confirm that the inclusion of the tide barely affects Sedna and 2012 VP$_{113}$, but find it does lower Leleakuhonua's primordial $\varpi$ by $\approx$$20^\circ$ (black triangle in Figure~\ref{fig:sednoids-rewind}), while producing a slightly larger uncertainty range (marked by gray triangles in Figure~\ref{fig:sednoids-rewind}) 4.5 Gyr ago.

We observe that
the galactic tide does not visibly modify $\varpi$ by imposing a different precession rate directly. The deviation between two nominal $\varpi$ evolutions occurred when Leleakuhonua's $q$ dropped to $\approx$$60$~au around $1$~Gyr ago (consistent with \citealt{Sheppard.2019}'s figure 7). A lower $q$ allows stronger diffusion in $a$ \citep{Hadden.2023}, which further disperses the $\varpi$ precession induced by the four giant planets (Eq.~\ref{eq:apsidal-precession-rate}).

In the presence of the  tide, the $p$ value of chance
primordial clustering is still $<$5$\%$. 
We find that the 
$\varpi$ uncertainty at $-4.5$~Gyr is currently dominated by the large $a$ uncertainty in Leleakuhonua's orbit (rather than whether or not the tide is included). 
We thus conclude that the primordial alignment in $\varpi$ remains an extremely interesting possibility, with (1) more sednoid orbits required to reach 3$\sigma$ confidence, and 
(2) high-precision orbits being needed for these large-$a$ objects.

\section{Cosmogonic Interpretations}\label{sub:interpertations}

Similar backward integrations have been successfully applied in identifying asteroid family. 
For example, \citet{Nesvorny.2002} used this method to identify 13 members of the Karin family, believed to have been created 
from a catastrophic collision of a larger parent body. 
All of its current presumed members have very similar $\Omega$ and $\varpi$ when propagated back to the breakup epoch 5.8 Myr ago. 

The sednoid story is similar, but not based on a collision.
Instead, sednoids (and other detached TNOs) are believed to have originated from the primordial planetesimal disk and were transplanted to their current orbits. 
If the primordial alignment is validated with other decoupled TNOs, it would mean that after formation the detached Kuiper Belt has remained largely unperturbed over the past $\approx$4~Gyr, with only precession induced
by the four giant planets slowly altering orbits. 
This would be incompatible with
hypotheses suggesting undiscovered planets currently existing in the outer Solar System (e.g., \citealt{Gladman.2002, Brunini.2002, Gomes.2006, Batygin.2016, Volk.2017, Lykawka.2023}), and aligns with the non-detection of planetary bodies in observational surveys \citep{Trujillo.2020, Belyakov.2022pj} and high-precision spacecraft tracking data \citep{Fienga.2020, Gomes.2023}.

This primordial alignment would strongly argue for a very early primordial event (within the first $\sim$100~Myr) that established this specific apsidal orientation for the strongly detached TNO populations, presumably at the same time when their $q$ were lifted. 
The exact timing and duration of the event depend on the width of the clustering, which is still uncertain due to 
the alignment being defined by only 3 objects.
Nevertheless, one can roughly estimate from Figure~\ref{fig:sednoids-rewind} that this event must have ended within a few hundred Myr, after which the clustering is no longer significant. 
One can thus potentially use the width and timing of the primordial alignment ($-4.5$--$-4.3$~Gyr) to constrain the sednoid perihelion lifting mechanism. We briefly explore two hypotheses that could explain this phenomenon.

\section{Possible Scenarios}\label{sub:scenarios}

\begin{figure*}[htb!]
  \centering
  \includegraphics[width=1.0\textwidth]{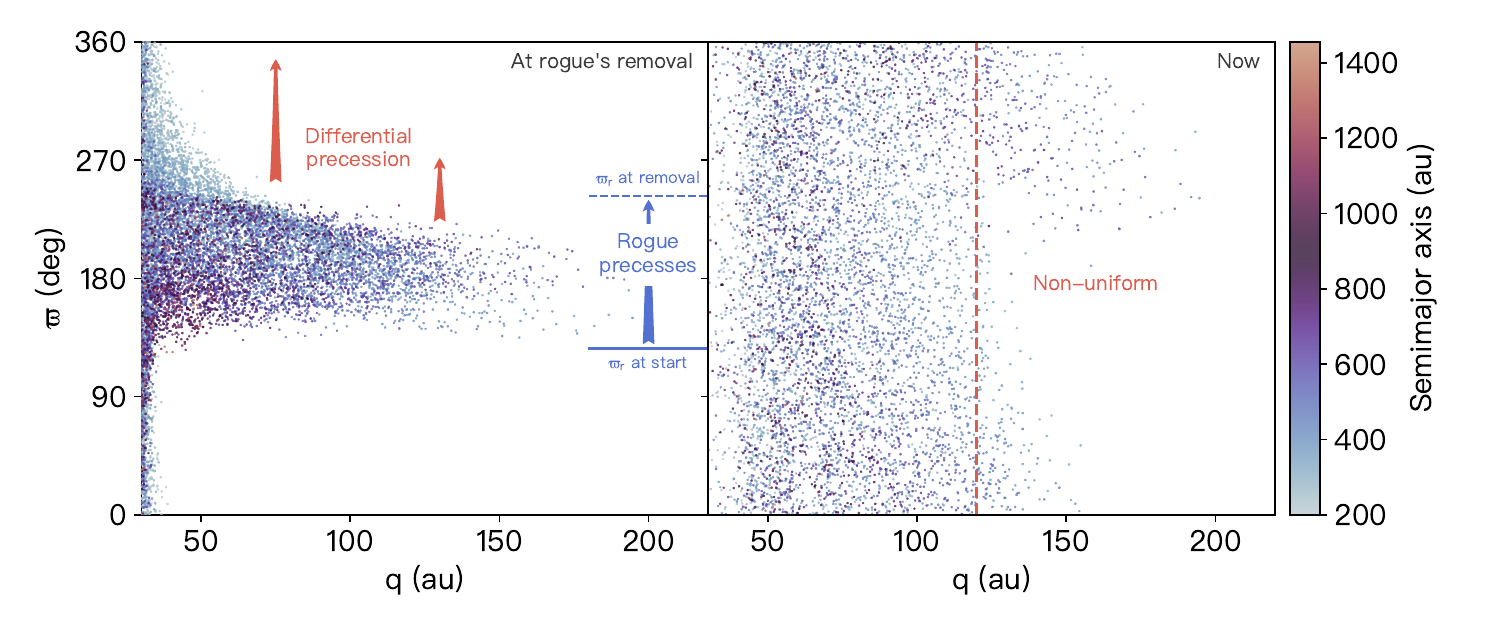}
\caption{TNO $\varpi$--$q$ distributions at the end of a 2~$M_\oplus$ rogue planet's 185-Myr early presence (left) and today (right). The rogue planet's longitudes of perihelion ($\varpi_r$) at the start and at the removal are marked by blue solid and dashed lines, respectively. Upon the rogue's removal, the differing precession rates due to wide ranges of $a$ and $q$ (red arrows; see also Equation~\ref{eq:apsidal-precession-rate}) led to the near homogenization of $\varpi$, except for $q > 120$~au TNOs (right of the red dashed line, right panel) where precession periods are comparable to the age of the Solar System.}

  \label{fig:varpi-distribution}
\end{figure*}

\subsection{Primordial Planet}
One hypothesis to explain the alignment revolves around a temporarily present rogue planet born in the Solar System. \citet{Gladman.2006} showed that scattering rogue objects could have raised TNO perihelia and created Sedna-like orbits; the secular $q$-lifting effect is dominated by the single most massive rogue. \citet{Huang.2022ml} recently demonstrated that the rogue can also help populate distant TNOs below $a < 100$~au by collaborating with Neptunian mean-motion resonances. Previous studies have shown that the detachment dynamics of a highly eccentric (rogue) planet is correlated with its relative apsidal orientation $\Delta \varpi$ \citep{Batygin.2017, Huang.2023t}. 
It is thus worth exploring 
a proof of concept that a rogue can produce
a primordial alignment.

Figure~\ref{fig:varpi-distribution} displays the simulation result of a rogue planet that lifts objects from a massive early scattering disk (which has $q < 40$~au). 
The 2~$M_\oplus$ rogue has an initial orbit $a_{r,0}=400$~au, $q_{r,0}=32$~au, $i_{r,0}=15^\circ$, and was propagated along with the four giant planets for 185~Myr. 
To generate the initial disk of scattering particles, Neptune was forced to migrate from 24 to 30~au through an outer planetesimal disk of 500,000 test particles spanning 24.5 to 33~au; the migration timescale and disk parameters are similar to those of grainy migration simulations \citep{Nesvorny.2016ij8, Nesvorny.2016}, but we find that these parameters play little role in the emplacement of detached TNOs beyond $a > 200$~au. 
As a result of Neptune scattering,
TNOs
were constantly fed to the large-$a$ region, where the rogue's gravity can raise the TNO perihelia \citep{Gladman.2006}. 
The planetary simulation (rogue scattering and Neptune migrating) was performed using \textsc{Rebound} \citep{Rein.2012}, while the test particle simulation was carried out by \textsc{Glisser}, a GPU N-body integrator based on \citet{Zhang.2022}.

During the $\sim$$100$-Myr temporary presence, the rogue was able to reproduce sednoids in a wide range of semimajor axes (a = 200--$\sim$1000~au, color-coded in Figure~\ref{fig:varpi-distribution}).
The detachment was strongest along the rogue's apsidal line \citep{Huang.2023t}, which precesses $\approx$$100^\circ$ 
(blue lines) until ejection by Neptune. 
This $2 M_\oplus$ rogue creates a strong clustering in the $\varpi-q$ space, with a circular standard deviation of $\approx$$25^\circ$ for $q > 50$~au particles. 
While this is more dispersed than the primordial alignment observed in the three Sednoids, more Sednoid discoveries will be needed to measure the underlying spread of their primordial apsidal lines. 
A tightly-dispersed primordial $\varpi$ distribution can occur for a more massive rogue planet with a shorter lifetime before ejection.

After the rogue's removal, the now-detached objects/sednoids precess at different rates (determined by their $a$, $q$, and $i$ as per Equation~\ref{eq:apsidal-precession-rate}). This eventually results in a nearly-uniform $\varpi$ distribution in today's surviving population. 
For icy bodies with $q > 120$~au, however, there is not enough time to homogenize their orbital orientations because their precession periods are comparable to the age of the Solar System. 
If this general picture is true, then if any $q > 120$~au TNOs were to be discovered, they would probably spread from $\varpi\approx$$-120^\circ(240^\circ)$ to $\approx$$60^\circ$ (Figure~\ref{fig:varpi-distribution}'s right panel), assuming a primordial $\varpi$ clustering near $200^\circ$.

The close primordial sednoid alignment is only created in $\varpi$, not in $\Omega$ or $\omega$; both angles have a larger circular standard deviation of $\approx$$80^\circ$ at 4.5 Gyr ago (but have an anti-correlation to produce the $\varpi$ clustering). 
This is also the case in the rogue planet simulation, where the detached TNOs possessed various $\Omega$ and $\omega$, instead of showing a strong clustered peak as in Figure~\ref{fig:varpi-distribution}'s left panel.

\begin{figure}[htb!]
  \centering
  \includegraphics[width=1.0\columnwidth]{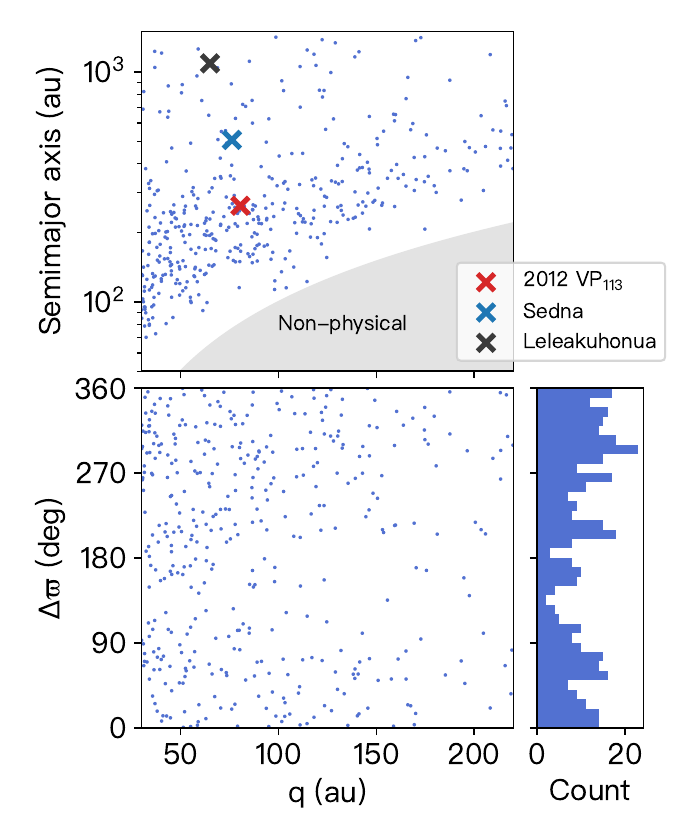}
\caption{Orbital distributions of TNOs detached by a solar-mass passing star ($q_\ast = 300$~au and $v_\infty=1$~km/s). The upper panel shows that this particular stellar encounter was able to create all three sednoids (crosses) in the $a$--$q$ space. However, there was no strong primordial clustering of $\varpi$ post flyby (lower panels). }
  \label{fig:passing-star}
\end{figure}

\subsection{Stellar Flyby}
Another scenario worth exploring is that the primordial alignment might correlate with a passing star's longitude of perihelion. \citet{Batygin.2020jr} analytically showed that the particle's eccentricity evolution under a passing star is related to their $\varpi$ difference. However, numerical studies of stellar encounters mainly focused on TNO $a$, $q$, $i$ distributions (e.g., \citealt{Ida.2000, Brasser.2012, Nesvorny.2023}) instead of their apsidal lines. Therefore, we conducted a simulation with a solar-mass star passing through a primordial scattering disk ($5 < q < 25$~au), with the closest stellar approach at $q_\ast = 300$~au \citep{Adams.2010, Batygin.2020jr}. Its hyperbolic trajectory has $i_\ast = 15^\circ$ relative to the Solar System and $v_\infty=1$~km/s, which is the typical velocity dispersion for young embedded clusters \citep{Brasser.2006}.

Figure~\ref{fig:passing-star} shows the result of this simulation.
Although such a passing star produced sednoids across a wide semimajor axis
range, no strong clustering was formed in the $\Delta \varpi$--$q$ panel.
One observes a weak preference for particles with $\Delta \varpi = 90^\circ$ and $270^\circ$ to be lifted from the scattering disk as the star passes (\ref{fig:passing-star}'s histogram) but this is insufficient to produce the correlation shown in Figure~\ref{fig:sednoids-rewind}. While this preliminary simulation does not favor the stellar encounter for primordially-clustered sednoids, further exploration of the star's parameters with a focus on the $\varpi$ distribution it creates is warranted.

\section{Discussion}\label{sec:dis}

\begin{figure*}[htb!]
  \centering
  \includegraphics[width=1.0\textwidth]{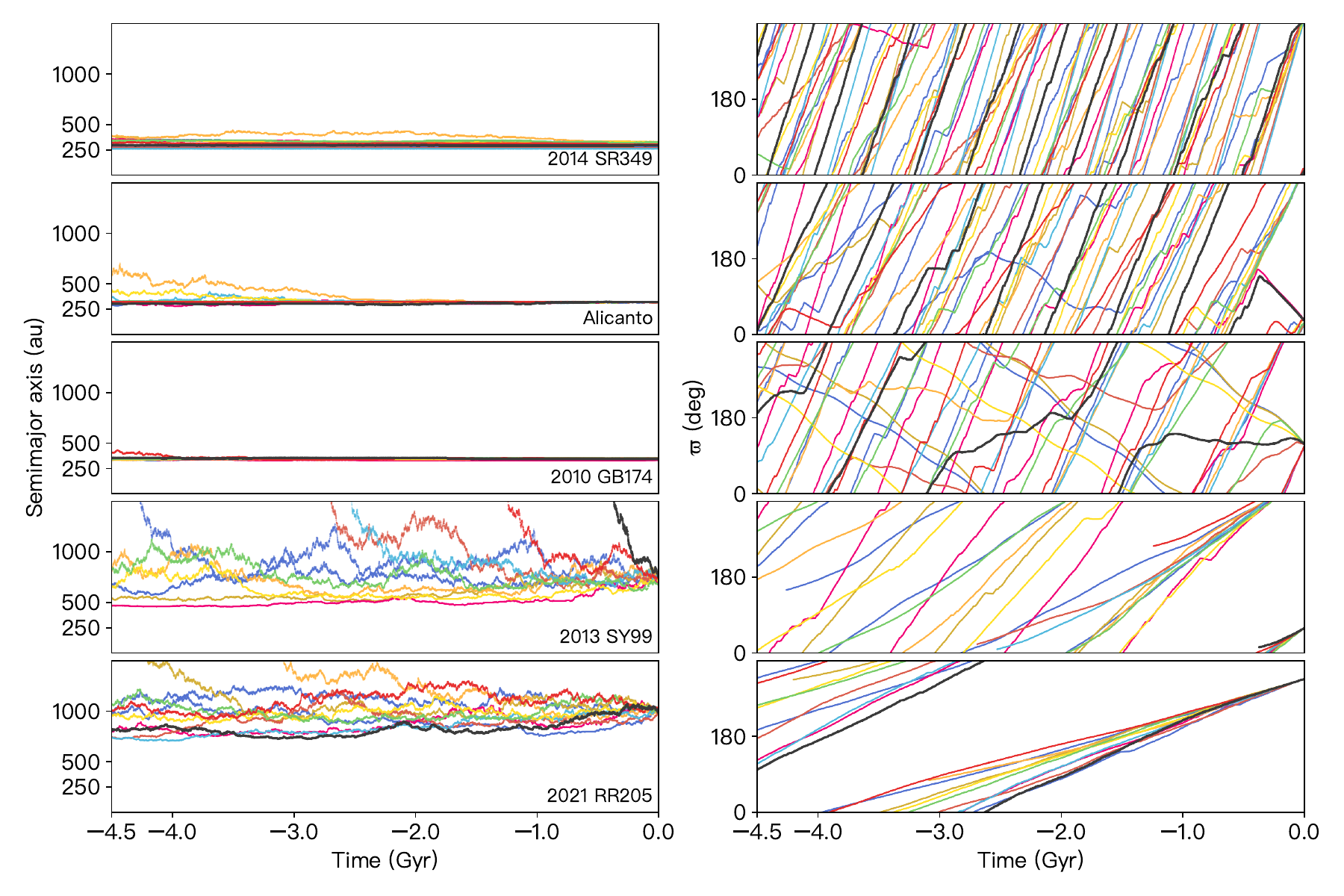}
\caption{Backward integrations of five large-$a$ and high-$q$ TNOs, taking into account their orbital uncertainties. Nominal orbits are colored black, while clones are in different colors. 
Although 2014 SR$_{349}$, Alicanto, and 2010 GB$_{174}$ show 
very similar 
$a$ evolutions 
across the clones (left panel), none of the integrated objects have 
sufficiently confined
$\varpi$ histories (right panel), rendering them still insufficient for the primordial alignment analysis.}
  \label{fig:other_tnos}
\end{figure*}

\textbf{Other candidates}. We also tried integrating backward another five large-$a$ and high-$q$ TNOs: 2014 SR$_{349}$, (474640) Alicanto (a.k.a. 2004 VN$_{112}$), 2010 GB$_{174}$,  2013 SY$_{99}$, and the recently-discovered 2021 RR$_{205}$. 
We judged these as the least prone to modifications by chaotic interactions with Neptune when the TNOs were near perihelion.

The $a$ and $\varpi$ histories are shown in Figure~\ref{fig:other_tnos}.
The first three objects with $a \approx 300$~au and $q\approx 45$~au have relatively stable semimajor axis histories. However, their $\varpi$ evolutions rapidly diverge from the start, with clones immediately possessing various rates of apsidal precession and even different directions; the latter is due to TNO proximity to very high-order Neptunian mean-motion resonances \citep{Volk.2022}. 
The last two objects, with $a \gtrsim 700$~au and $q \gtrsim 50$~au, are slowly scattering in $a$ due to dynamical diffusion caused by Neptune resonance overlap \citep{Batygin.202150u, Hadden.2023}, with some clones not even stable over $\sim$4~Gyr. This results in a wide range of precession rates $\dot{\varpi}$, leading to indeterminate primordial $\varpi$. For 2014 SR$_{349}$, 2013 SY$_{99}$, and 2021 RR$_{205}$, future observations could potentially improve the uncertainties to the degree where a sufficiently-precise determination of their primordial $\varpi$ may be possible.

\vspace*{-3mm}

\section{Acknowledgements}

We thank W.~Fraser, J.~Kavelaars, R.~Pike, D.~Raggozine, and K.~Volk for useful discussions, and an anonymous referee for providing insightful comments. Y.H. is grateful to W.~Zhu and Tsinghua University for their generous support. B.G. acknowledges Canadian funding support from NSERC. 

\bibliography{_ref_fixed}{}
\bibliographystyle{aasjournal}

\end{document}